\documentclass[pra,reprint,showpacs,showkeys,groupedaddress,nofootinbib,floatfix]{revtex4-1}
\usepackage{amsmath,amssymb,amsfonts}
\usepackage{graphicx}
\usepackage{txfonts}
\usepackage{epstopdf}
\usepackage[T1]{fontenc}
\usepackage{color}

\usepackage[unicode]{hyperref}
\hypersetup{
    unicode=true,                                           
    a4paper=true,
    plainpages=false,
    pdffitwindow=true,                                      
    colorlinks=true,                                        
    linkcolor=blue,                                        
    citecolor=blue,                                         
    filecolor=blue,                                        
    urlcolor=blue                                          
}
\newcommand{\bra}[1]{\langle #1\vert}
\newcommand{\ket}[1]{\vert #1\rangle}
\newcommand{\bket}[2]{\langle #1\vert#2\rangle}
\newcommand{\ev}[1]{\langle #1 \rangle}

\newcommand{\pdsq}[1]{\frac{\partial^2}{\partial #1^2}}

\begin{document}

\title{Phase-matching condition for enhanced entanglement of colliding indistinguishable quantum bright solitons in a harmonic trap}
\author{David I. H. Holdaway}
\email{David.Holdaway@dunelm.org.uk}
\author{Christoph Weiss}
\author{Simon A. Gardiner}

  \affiliation{Joint Quantum Centre (JQC) Durham--Newcastle, Department of Physics, Durham University, Durham DH1 3LE, United Kingdom}

\date {\today}

 \begin{abstract}
We investigate finite number effects in collisions between two states of an initially well defined number of identical bosons with attractive contact interactions, oscillating in the presence of harmonic confinement in one dimension.  We investigate two $N/2$ atom bound states, which are initially displaced (symmetrically) from the trap center, and then left to freely evolve.  For sufficiently attractive interactions, these bound states are like those found through use of the Bethe Ansatz (quantum solitons). However, unlike the free case, the integrability is lost due to confinement, and collisions can cause mixing into different bound state configurations.  We study the system numerically for the simplest case of $N=4$, via an exact diagonalization of the Hamiltonian within a finite basis, investigating left/right number uncertainty as our primary measure of entanglement.  We find that for certain interaction strengths, a phase matching condition leads to resonant transfer to different bound state configurations with highly non-Poissonian relative number statistics.
 \end{abstract}


\maketitle

\section{Introduction}

Bright matter-wave soliton solutions have been predicted in the attractive (self focusing) one dimensional non-linear Schr\"odinger equation (1D NSLE) for some time~\cite{ZakharovShabat1974}.  Soliton are self localizing wavepackets, where collisions with other solitons do not change the  asymptotic shape, speed or amplitude (number of atoms) in either soliton. Only the position and phase are modified from what they otherwise would have been~\cite{AblowitzSegurBook1981}. The 1D NLSE is known to describe the dynamics (at a mean-field level) of an ultracold Bose gas in tight radial confinement, such that the radial degree ($y$ and $z$) of freedom are in the ground state of the potential and essentially decouple from the dynamics in the unconfined ($x$) dimension. The parameter regimes required for this decomposition are discussed in the next section.  Outside of this 1D regime, attractive condensates do not necessarily have a local potential minimum present at a finite width (and thus a metastable ground state). If no metastable state exists, the gas will undergo an $s$-wave collapse, observed by several groups experimentally~\cite{SackettHulet1999,GertonHulet2000,CornishThompson2006}, which limits the number of atoms considerably.  

Recent experiments~\cite{MarchantCornish2013,MinarThesis,DuarteHulet2011,Hulet2010b,CornishThompson2006,StreckerHulet2003,Cornish2003} have produced condensates with attractive interactions, and observed self-localized wavepackets.  Due to the presence of confinement in the axial ($x$) direction, and in some cases insufficient fulfillment of the conditions ncessary for effective 1D behaviour, these states were not solitons in the strictest sense.  External potentials in the $x$ direction and residual 3D effects break the integrability of the system, and without this property soliton collisions can transfer some kinetic energy into breathing modes~\cite{BillamCornish2011} or generate entanglement~\cite{LewensteinMalomed2009}.  Additionally, there have been questions as to whether and how long coherence was present in trains of multiple solitons, with some papers looking beyond the mean field models predicting coherence to be short lived~\cite{Streltsov2011}, and the presence of modulational instabilities~\cite{CarrBrand2004}.  Current experiments report some progress towards seeing splitting in a ``fast scattering'' regime~\cite{Hulet2010b,MarchantCornish2013}; other bright-soliton experiments in Bose-Einstein condensates can be found in Refs.~\cite{KhaykovichSchreck2002, StreckerPartridge2002,EiermannOberthaler2004}.

In order to address the issue of macroscopic coherence it is necessary to move beyond the mean field theory.  We model the $N$-particle quantum dynamics by assuming a zero-range pseudo potential, valid when the spacing between particles is much less than the characteristic distance of the true interaction potential between each of the atoms.  This model has been shown to be a valid reduction from the 3D many-body theory (given the conditions we discuss in Sec.~\ref{sec:Hamandrescale}), at least in the case of repulsive gases~\cite{SeiringerYin2008}, and is commonly referred to as the Lieb--Liniger (LL) model~\cite{LiebLiniger1963}.  If no external potentials are present in the $x$ dimension, the LL model can be solved analytically and provide a many-body level description of the system. Bound state solutions also exist within the (unconfined) LL model with attractive interactions, which we will refer to as quantum solitons; the number of atoms in each of these bound states is a good quantum number and so cannot change as solitons collide or move. In the limit of large number the asymptotic position and phase shift from the mean field model are recovered~\cite{LaiHaus1989B}.  The eigenstates in this system are completely delocalized in position (reflecting the translational invariance of the Hamiltonian), however the quantum solitons have been shown to have the same density~\cite{CastinHerzog2001} about the center-of-mass as mean field solitons, and even very similar many-body wave functions in their internal degrees of freedom~\cite{HoldawayWeiss2012}.  Proposals for squeezing~\cite{CarterEtAl1987} and non-demolition measurement~\cite{DrummondEtAl1993} have also been suggested, in the context of optical fibers, along with Anderson localization~\cite{DelandeEtAl2013}.

In this work we investigate the LL model, with the addition of a harmonic confinement term, starting with two identical quantum solitons, equally displaced from the trap center and left to freely evolve.  Similar to the mean-field case, the confinement breaks the integrability of the system; while there is no universally agreed definition of quantum integrability~\cite{SutherlandBook2004}, this would break any such condition.  In the regime where the harmonic oscillator length is much larger than the soliton size, this term is essentially perturbing a system of solitons. Mixing between different bound states is now possible during collisions, as is entanglement generation between colliding quantum solitons.  We have investigated this effect for repulsive and weakly attractive gases in our previous work~\cite{HoldawayWeiss2013}, but qualitative differences emerge in the more strongly attractive case. 
This effect has been examined for colliding distinguishable solitons~\cite{GertjerenkenEtAl2013}; in this system, generation of mesoscopic Bell states via the scattering of distinguishable bright solitons would even be possible without harmonic confinement. We note additional terms which break integrability have also been considered~\cite{LewensteinMalomed2009,GarnierAbdullaev2006}, including narrow barriers, leading to mean-field level splitting for fast soliton-barrier collisions~\cite{GarnierAbdullaev2006,Helm2012} and center-of-mass wave function splitting for sufficiently slow collisions~\cite{GertjerenkenWeiss2012scattering}.  Current experiments report some progress towards seeing splitting in a ``fast scattering'' regime~\cite{Hulet2010b,MarchantCornish2013}; other bright-soliton experiments in Bose-Einstein condensates can be found in Refs.~\cite{KhaykovichSchreck2002, StreckerPartridge2002,EiermannOberthaler2004}.

The main focus of this work is a surprising resonant transfer of two initially independent but indistinguishable bright solitons into a quantum superposition of $N-1$ particles being on the right and one particle on the left and vice versa. This highly non-classical quantum superposition is similar to a Schr\"odinger-cat state (which has also been called NOON-state~\cite{WildfeuerDowling2007}). Non-classical quantum superpositions such as Schr\"odinger-cat states are relevant for quantum-enhanced interferometry~\cite{GiovannettiEtAl2004}.  While it might sound tempting to realize such a state as, say, the ground state of an attractively interacting Bose-Einstein condensate in a double well, such an approach will not be successful in the presence of tiny asymmetries~(cf.~\cite{WeissTeichmann2007}) and decoherence.  In order to minimize effects of decoherence, mesoscopic quantum superpositions in Bose-Einstein condensates will ideally be realized dynamically on short time scales (Refs.~\cite{MicheliEtAl2003,MahmudEtAl2005,WeissCastin2009,DagninoEtAl2009,StreltsovAlon2009,GarciaMarchEtAl2011,MazzarellaEtAl2011} and references therein),  however we do not consider decoherence in this work.  Furthermore, dynamic realizations of mesoscopic quantum superpositions can be more robust to asymmetries~\cite{WeissTeichmann2007}; in our case asymmetries in the initial condition primarily lead to breathing modes of the total center-of-mass wave function without affecting the entanglement production (Sec.~\ref{sec:init_cond}). 

The paper is organized as follows:  Section~\ref{sec:sys} introduces a quasi-analytic model of the system and the dimensionless length and time rescaling that is used throughout the paper.  Section~\ref{C4:singlet_trimer_mix} introduces a method based on time dependent perturbation theory to study transfer between different number configurations and predict interaction strengths which give resonant transfer.  Section~\ref{sec:numerical} discusses the numerical method, based on exact diagonalization, and the operator expectation values which we study.  Section~\ref{sec:results} presents numerically obtained results for the evolution of our observables and entanglement measures in the system, and compares the predictions of our perturbation theory with the numerical results.  Section~\ref{sec:conc} summarizes the main conclusions and highlights an outlook for future research into this system.

\section{Quasi-analytical model \label{sec:sys}}

\subsection{Hamiltonian and rescalings \label{sec:Hamandrescale}} 
The model system we consider is an ultracold gas of bosonic atoms, held within a strongly anisotropic harmonic potential $V= m[\omega_{x}^{2}x^{2} + \omega_{\perp}^{2}(y^{2}+z^{2})]/2$, where $m$ is the atomic mass, and $\omega_{x}$, $\omega_{\perp}$ are the axial and radial harmonic trapping frequencies.  The radial degrees of freedom are assumed to be in the ground state, and so remain stationary throughout any time evolution in the x direction.  Sufficient conditions for the validity of such a 1D description are that the temperature $T$ satisfies the inequality $k_B T < \hbar \omega_{\bot}$, and the radial harmonic oscillator length satisfies\footnote{This condition is valid so long as the atomic density does not exceed that of the ground state in the absence of trapping in the $x$ direction. A more general condition would be $N|a_s||\psi(x)|^2 \ll 1$ with $|\psi(x)|^2$ the atomic density in the $x$ direction, normalized to unity.} $N|a_s| \ll a_{\bot}$,~\cite{SalasnichParola2002} where $k_{\mathrm{B}}$ is Boltzmann's constant, $N$ is the particle number, and $a_{s}$ is the $s$ -wave scattering length. Interactions between bosons are assumed to occur over distances much less than the average inter-particle spacing, which allows for the use of a zero-range pseudo potential, so long as it reproduces the $s$-wave scattering length of the true scattering potential.  We therefore consider a fully quantum mechanical Lieb--Liniger model Hamiltonian~\cite{LiebLiniger1963} for $N$ bosons in a 1D configuration, with the addition of a harmonic trapping potential.

We rescale to harmonic oscillator units (codified as $\hbar = \omega_{x} = m =1$, where $\omega_{x}$ is the axial harmonic trapping frequency and $m$ is the atomic mass), meaning that length is in units of $\sqrt{\hbar/m\omega_{x}}$, time in units of $1/\omega_{x}$, and energy in units of $\hbar \omega_{x}$; a harmonic oscillator period is then $2 \pi$.  Our many-body Hamiltonian is therefore given by 
\begin{equation}
H(x_{1},\ldots,x_{N}) = \sum_{k=1}^N \left( -\frac{1}{2} \pdsq{x_{k}} + \frac{x_{k}^2 }{2} \right)
 + g \sum_{k=2}^{N} \sum_{j =1}^{k-1} \delta(x_{k}-x_{j}) \; ,
\label{Ham:first_quant_ham}
\end{equation}
where $g=2 \hbar \omega_{\bot} a_s \sqrt{m/\hbar^3 \omega_x}$ is a dimensionless coupling parameter. For the extent of this paper we will have $g<0$, corresponding to attractive interactions.  If the confinement term is ignored, this Hamiltonian would be exactly diagonalizable, with all the eigenstates given by the Bethe Ansatz~\cite{Dorlas1993}.  These states have asymptotic momenta as good quantum numbers (which can be specific complex values in bound states). Naturally, the momentum cannot be a good quantum number with confinement present and so these states would no longer be eigenstates.  A strongly correlated 1D tunable Bose gas has been achieved with Cesium atoms~\cite{Haller2009}; other experiments using $^{85}$Rb~\cite{MarchantCornish2013} and $^7$Li~\cite{GertonHulet2000,MinarThesis} with attractive interactions have also been performed in this 1D regime.  
We note that in the limit $g \to \infty$, the eigenstates of the system can again be solved analytically~\cite{YukalovGirardeau2005} as the states map from those of a non interacting, spin-polarized Fermi-gas~\cite{MinguzziGangardt2005}.  Such a system has also been realized experimentally~\cite{Kinoshita2006}.

\subsection{Initial conditions \label{sec:init_cond} }

As is the case in~\cite{HoldawayWeiss2013}, we consider an initial condition constructed by taking two $N/2$-atom ground states (for a given $g$), equally and oppositely displaced from the trap center by a distance $x_0$, and symmetrizing the state. The initial ($t=0$) wave function is then
\begin{multline}
 \psi(x_{1},\ldots,x_{N},0) = \frac{B}{\sqrt{N!}} \sum_{\{\cal P \}}  f^{(N/2)}(x_1+x_0,\ldots,x_{N/2}+x_0) \\
 \times f^{(N/2)}(x_{N/2+1}-x_0,\ldots,x_N-x_0) \;,
\label{eq:initcon2}
\end{multline}
where $\{{\cal P }\}$ is the set of all permutations of $x_{1},\ldots,x_{N}$, $B$ is a normalizing factor and $f^{(N/2)}(x_1,\ldots,x_{N/2})$ is the ground state for $N/2$ interacting atoms in the harmonic trap.  This initial state is well motivated for $x_0 \gtrapprox 2$ (which is the case for all numerics in this paper), where the left and right sides have negligible initial overlap, but always constitutes a valid many-body wave function.  We also use a low-temperature theory which implies that each soliton (prepared separately) is in the ground state, and thus have individual center-of-mass wave functions which are Gaussian. A suggestion for how this initial condition could be realised in an experiment with ultracold atoms was given in~\cite{HoldawayWeiss2013}. Briefly summarized this involved using an optical super lattice~\cite{Sebby-Strabley2006}, with exactly $N/2$ attractive atoms per lattice site,\footnote{Having exactly $N/2$ atoms per site, as opposed to a delocalized superfluid state, could be achieved by starting with a MOT insulator regime~\cite{GreinerBloch2009}, increasing the lattice site depth and then adiabatically tuning the interactions to be the correct negative scattering length.} with the sites too deep to allow tunneling between them. The number of lattice sites is then halved by reducing the power from the high frequency beam, reducing the super-lattice to an ordinary lattice.  The states would then sit in a new wider potential, which is approximately harmonic. The wave function $\psi(x_{1},\ldots,x_{N},0)$ is therefore an experimentally realistic initial state, given sufficiently low temperatures.

Conveniently the global center-of-mass wave function of $\psi(x_{1},\ldots,x_{N},0)$ is in the ground state~\cite{HoldawayWeiss2013}, which will be the case for all time as the center-of-mass component of the Hamiltonian \eqref{Ham:first_quant_ham} separates off and commutes with the rest of the Hamiltonian. If we were to consider some slight asymmetry with the states not equidistant from the trap center (say one located at $x_0 + x'$ and the other at $-x_0+x'$), the center-of-mass wave function would simply oscillate periodically with an amplitude of $x'$ without changing the relative degrees of freedom.  Likewise if $f^{(N/2)}(x_{N/2+1}-x_0,\ldots,x_N-x_0)$ were $N/2$ body ground states of a slightly different harmonic potential to the final one (which might be somewhat inevitable with the lattice scheme we suggest) the centre-of-mass wave function would periodically breathe.  The relative degrees of freedom would be in a slightly excited state, but in the limit of $N g/2 \ll -1$ this change would be minor as the length scale of the bound states is set mainly by the interactions.  As a result this initial condition is not significantly affected by slight changes to these conditions.

The two body case $f^{(2)}(x_1,x_2)$ is known analytically~\cite{BuschEnglert1998,SowinskiBrewczyk2010}, but for larger $N$ these states must to be determined numerically.  However, in the limit $N|g| \gg 1$ and $g<0$ we can use as an ansatz the free (no external potential) ground state solution for the relative coordinates with the center-of-mass component in Gaussian profile such that this degree of freedom is in the trap ground state:
\begin{equation}
\begin{split}
f^{(n)} \sim &\sqrt{\frac{|g|^{N-1} (N-1)!}{\sqrt{N \pi}}} \exp\left(-\frac{(x_1 + .. + x_n)^2}{2n}\right)  \\
&\times \prod_{1 \le j < k \le n} \exp\left( -\frac{|g|}{2} \vert x_k-x_j \vert\right) \;. 
\end{split}
\end{equation}
The energy of such a state, placed into a harmonic potential, is known analytically, and is given (in units with $\hbar \omega_{x}=1$) by~\cite{HoldawayWeiss2012}
\begin{equation}
E^{(n)} =  \frac{1}{2} -\frac{g^2 n (n+1)(n-1)}{24} + \sum_{k=1}^{n-1} \frac{1}{k^2 g^2 n} \;.  
\label{eq:natomenergy}
\end{equation}
The $1/2$ term is from the Gaussian envelope, the second term is the free soliton energy and the third is the first order correction from the confining potential.  Additionally, variational techniques can be used to better estimate the wave function and energy~\cite{HoldawayWeiss2012}.

\subsection{Left-right separation of the Hamiltonian} 

Without interactions, our two clusters would simply oscillate with a period of $\pi$ (due to the left-right symmetry the periodicity is halved).  Interactions between the left and right clusters break this periodicity, complicating the dynamics.  Likewise, with interactions but no confinement a collision between two bound states can only result in the same two bound states emerging, with only an asymptotic position and phase shift\footnote{Relative phase between two states of definite number is actually not defined, but when the incoming states are superpositions of number states a phase difference can be defined~\cite{Peggbarnett1997}.}. The addition of confinement allows collisions to mix the wave function into states which are a superposition of two separate $n$ and $N-n$ body clusters, to the left and right respectively (in order to preserve the left-right symmetry of the state).  It is also possible to mix into states with more than two bound state clusters, however for simplicity we temporarily neglect this effect in order to treat the situation analytically.  

We are interested in how such states evolve in time and whether it is possible to predict quantities like the single body density and conditional position expectation values following a measurement of the number left and right of the center.  The symmetrization of the available states which can be mixed to be a transfer interaction is important only when the two clusters are not well separated.  Assuming they are well separated, we can consider the evolution separately by splitting the $N$ body Hamiltonian \eqref{Ham:first_quant_ham} within the region $x_1 \le x_2 \cdots \le x_N$ (sufficient by Bose symmetry) into two separate $n$ and $N-n$ body Hamiltonians $H_L$ and $H_R$, and an interaction term $H_I$:
\begin{align}
 H_L(n) &= \sum_{k=1}^n \left( -\frac{1}{2} \pdsq{x_{k}} + \frac{x_{k}^2 }{2} \right) + g \sum_{k=2}^{n}   \delta(x_{k}-x_{k-1})  \; , \nonumber \\
 H_R(n) &= \sum_{k=n+1}^N \left( -\frac{1}{2} \pdsq{x_{k}} + \frac{x_{k}^2 }{2} \right) + g \sum_{k=n+2}^{N} \delta(x_{k}-x_{k-1})  \; , \nonumber \\
 H_I(n) &= g \delta(x_{n+1}-x_{n})  \; .
 \label{eq:split_ham_2}
\end{align}
Due to the region we have restricted the Hamiltonian to, only one term in $H_I(n) = g[\delta(x_{N}-x_{1}) +\cdots+ \delta(x_{n+1}-x_{n})]$ is ever non zero, which is the only one we include. We have $[ H_L(n),H_R(n) ] = 0$ and hence we can combine eigenstates of each Hamiltonian to create eigenstates of $H_L(n)+H_R(n)$. Noting that we can neglect $H_I(n)$ when the clusters are far from the trap center, $H_L(n)+H_R(n)$ can there be considered to be the total Hamiltonian when the states are not undergoing a collision.  Each Hamiltonian $H_L(n)$ can again be split into a center-of-mass $H_{L/R}^{(\rm C)}$ and relative part $H_{L/R}^{(\rm r)}$, which again commute, (note for $n=1$ there is only the center of mass).  
The center-of-mass coordinate of the left side is $x_{\rm C} = (x_1 + \cdots +x_n)/n$ and so we have
\begin{align}
H_{L}^{(\rm C)} = \frac{1}{2} \left( -\frac{1}{n}\pdsq{x_{\rm C}} + n x_{\rm C}^2\right) \;.
\end{align}

\subsection{Construction of oscillating quantum soliton states \label{sec:cohstatemod}} 
In order to make analytical predictions, we consider states which are $n$ atom clusters, with center-of-mass wave function that remains Gaussian throughout its evolution (with a phase gradient and changing position expectation value) displaced from the center by $X_n$ at $t=0$, and thus with a potential energy of $(nX_n^2+1)/2$.  For the relative degrees of freedom we will assume each side is in the {\it relative ground state} for $n$ attractive atoms, with eigenenergies $E^{(n)}$ strictly less than the $g \to -\infty$ limit given by Eq.~\eqref{eq:natomenergy}.  For $n=2$ the exact energy and the limiting values are correct to one percent when $g<-2.3$.  Taking a left state, we can consider the time evolution far from the center (where interaction terms and symmetry become important) to be described by
\begin{align} 
\psi_{n}(x_1,\ldots,x_n,t) = \exp\left[ -i t H_L(n) \right] \psi_{n}^{X_n}(x_1,\ldots,x_n,0) \;.
\label{eq:time_evolution}
\end{align} 
The relative degrees of freedom are in the ground state and so evolve only in phase, and the center-of-mass undergoes simple harmonic motion (see for example~\cite{MartinAdams2008}).  For convenience will now drop the coordinate notation in the many-body wave functions.

We must also have a right cluster of $N-n$ atoms, initially located at $-X_{n} n/(N-n)$, due to the constraint that the global center-of-mass position expectation value is located at $x=0$.  Adding the left and right center-of-mass energies $\ev{H_{L}^{(\rm C)}}+\ev{H_{R}^{(\rm C)}}$,\footnote{The  expectation value notation $\ev{\;}$  is defined here via $\ev{O(x_1,\ldots,x_N)} = \int_{-\infty}^{\infty} dx_1 \ldots \int_{-\infty}^{\infty} dx_N \psi^*(x_1,\ldots,x_N) O(x_1,\ldots,x_N) \psi(x_1,\ldots,x_N)$, but due to the definition of the Hamiltonians only on $x_1 \le x_2 \cdots \le x_N$ these are best evaluated as $\ev{H(x_1,\ldots,x_n)} = N!\int_{-\infty}^{x_2} dx_1 \int_{-\infty}^{x_3} dx_3 \ldots \int_{-\infty}^{\infty} dx_N \psi^*(x_1,\ldots,x_N) H(x_1,\ldots,x_n) \psi(x_1,\ldots,x_N)$, with the $N!$ term dealing with identical permutations.} the total energy of such a state is given by
\begin{align}
E_{n}+E_{N-n} \sim -\frac{N g^2}{24}\left[ N^2 - 3n(N-n) -1\right] +  \frac{N n}{N-n} X_{n}^2 + 1 \;.
	  \label{eq:nN-nenergy}
\end{align}

We define $\psi_{n,N-n}(0)$ to be an $n$ atom cluster to the left and an $N-n$ cluster to the right, displaced by $X_{n}$ and  $-X_{n} n/(N-n)$ respectively (with unit norm), such that our initial condition is $\psi_{N/2,N/2}(0)$.  When left and right are well separated, the time evolution is determined by the first two terms in Eq.~\eqref{eq:split_ham_2}
\begin{align} 
\psi_{n,N-n}(t)  = &\exp\left[ -it H_L(n) \right] \psi_{n}^{(X_n)}(0) \nonumber \\
&\otimes\exp\left[ -it H_R(n) \right]  \psi_{N-n}^{(-nX_n/[N-n])}(0) \;.
\label{eq:time_evolution_coherent}
\end{align} 
All possible wave functions that can evolve from a symmetric initial condition must also posses left-right symmetry, and so wave functions must be a {\it symmetrized} product of both sides.  These wave functions are also implicitly Bose symmetric, as it is only defined on the fundamental region $x_1 \le  x_2 \le ... \le x_N$. This fact is not mathematically significant when the states are well separated, but becomes important during collisions. We will treat interactions during collisions perturbatively later in Sec.~\ref{C4:singlet_trimer_mix}, using these eigenstates of $H_L(n)$ and $H_R(n)$. Hence we define the left and right symmetric state 
 \begin{align} 
 \varphi_{n,N-n}(t) = 
  \begin{cases}   {\displaystyle {\cal N}_{n,N-n}(t) \left[\psi_{n,N-n}(t) +  \psi_{N-n,n}(t) \right]  \:\mbox{if}\: n \ne \frac{N}{2}\;,  } 
  \\
{\displaystyle		 \psi_{N/2,N/2}(t) \quad \mbox{if}\quad n = \frac{N}{2}\;,}
  \end{cases}
 \label{eq:possible_wf}
\end{align} 
 with the normalization term
\begin{multline} 
\left[{\cal N}_{n,N-n}(t)\right]^{-2} 
= N!
\int_{-\infty}^{x_2} dx_1 \int_{-\infty}^{x_3} dx_2 \ldots  
\\ \ldots  \int_{-\infty}^{\infty} dx_N \vert \psi_{n,N-n}(t) + \psi_{N-n,n}(t) \vert^2 \;.
\end{multline} 
The strange integration range and prefactor of $N!$ are due to the definition only on the simplex region $x_1 < \ldots < x_N$.   
If we assume the centres of Gaussian wavepackets describing the centres of each of the states to be separated by some distance $X$, then if the states are well separated ($X \gg 1$) for $n' \ne n$ and $n' \ne N$ we have $\bket{\psi_{N-n,n}(t)}{\psi_{N-n',n'}(t)} \sim \exp(-X^2)$, and so ${\cal N}_{n,N-n}(t) \to 1/\sqrt{2}$ as $X \to \infty$.  For finite separations this normalization term may be different, but its exact value is not currently important.

Reintroducing interactions, each $H_{I}(n)$ from Eq.~\eqref{eq:split_ham_2} can mix different $\ket{\varphi_{n,N-n}(t)}$. It will also affect the relative position between the two sides, an effect that was explored significantly in our previous work~\cite{HoldawayWeiss2013}.  We temporarily neglect this effect for the purpose of this analysis, but note it will introduce a greater uncertainty in positions at late times. Within this set of approximations, we can express any possible wave function the system can take as
\begin{align} 
\psi(t) &\simeq \sum_{n=1}^{N/2} c_{n,N-n}(t) \varphi_{n,N-n}(t) \;.
 \label{eq:possible_wf2}
\end{align} 
The single cluster state is assumed to have a negligible contribution due to reasons of energy and center-of-mass momentum conservation.  The coefficients $c_{n,N-n}$ are those considered in Sec.~\ref{C4:delta_int_pert}. In the case of our four atom system, these approximations give us the simple wave function
\begin{align} 
 \psi(t) \simeq c_{2,2}(t) \varphi_{2,2}(t) + c_{1,3}  \varphi_{1,3}(t)\; ,
 \label{eq:N=4_pos_wf}
\end{align} 
with the initial condition that $c_{2,2}(0) = 1$.  To zeroth order in $H_I$ the dimers would simply oscillate perfectly with a period of $\pi$, this is in principle obtained in the limit the initial separation tends to infinity, or trivially when $g \to 0$.  The mixing between the $N/2,N/2$ and $n,N-n \; + \; N-n,n$ states can equivalently be seen as being due to the coupling between the Bethe ansatz eigenstates due to the harmonic trapping, or from the $H_I(n)$ in our coherent state model.  The rate of transfer should depend on this coupling, which we use as a parameter in Sec.~\ref{C4:singlet_trimer_mix}, but both coupling terms are difficult to calculate directly.

\subsection{Predictions of oscillation amplitudes}
If we assume our state remains of the form Eq.~\eqref{eq:possible_wf2}, we can make analytic predictions of $X_{n}$, the maximum displacements of the oscillating clusters in $\varphi_{n,N-n}(t)$, by assuming they have the same energy as $\varphi_{N/2,N/2}(t)$.  This condition $E_{N/2,N/2} = E_{n,N-n}$ with $E$ defined in Eq.~\eqref{eq:nN-nenergy}, implies
\begin{align}
&N x_0^2 + 2 E^{(N/2)}  = \frac{N n}{N-n} X_n^2 +E^{(N-n)} + E^{(n)} \;,
\end{align}
with $E^{(n)}$ given in Eq.~\eqref{eq:natomenergy} and $x_0$ the initial position of the $N/2$ clusters.  Within the strongly interacting regime, one can neglect contributions of order $1/g^2$ and simplify this expression to
\begin{align}
X_n^2 =  \frac{N-n}{n} \left\{ x_0^2 + \frac{g^2}{8} \left[ \frac{N^2}{4} - n(N-n) \right] \right\}\;,
\label{eq:E1=E2}
\end{align}

To estimate the uncertainty in these values due to the possibility of collisions mixing to states with a different energy, we can derive bounds based on the Hamiltonian variance.  These bounds should be considered weak and subject to all the prior assumptions in Sec.\ref{sec:cohstatemod}.  As our Hamiltonian is time independent, the variance of its expectation value 
\begin{align}
 \Delta E &\equiv \sqrt{\ev{H^2} - \ev{H}^2 } \;,
\label{eq:DeltaE}
\end{align}
 is constant. This is because the time evolution operator $U(t)=\exp(-iHt/\hbar)$ [with $|\psi(t)\rangle = U(t)|\psi(0)\rangle$] commutes with $H^{\nu}$ ($\nu=1,2,3,\ldots$), hence we have: $\langle\psi(t)| H^{\nu}|\psi(t)\rangle =\langle\psi(0)| H^{\nu}|\psi(0)\rangle $ for and positive integer $\nu$. This remains true if one shifts $H$ by a constant offset value.  We consider only the $N=4$ case as this is used in the numerics.  
 
 In the limit $x_0 \gg 1$, we can treat each side separately to get an analytic expression~\cite{HoldawayWeiss2013}
\begin{align}
 \Delta E \to 2 x_0  \;.
\label{eq:DeltaEappx}
\end{align}
When the two states in our model have negligible overlap, one can derive the bound on the energy difference (c.f.~appendix in~\cite{HoldawayWeiss2013})
\begin{equation}
 \vert E_{2,2} - E_{3,1} \vert \le  \frac{\Delta E}{\sqrt{p (1-p)}} \; ,
\label{eq:Ediff_ineq}
\end{equation}
 with $p= |c_{2,2}|^2$. This bound tends to infinity as $p$ tends to $1$ or $0$, but this would imply there is no occupation of one of the states anyway, so this is physical.  This modifies Eq.~\eqref{eq:E1=E2} to an inequality, so that in the strongly interacting limit we have
\begin{equation}
\left(\frac{g^2}{8}  + x_0^2  - \frac{\Delta E}{2\sqrt{p(1-p)}} \right) \le \frac{X_{1}^2}{3} \le  \left(\frac{g^2}{8}  + x_0^2 +\frac{\Delta E}{2\sqrt{p(1-p)}}  \right) \; .
\label{eq:coherentbounds}
\end{equation}
This allows for additional discretion in the particles kinetic energy and thus maximum position reached after each collision.  The upper bound is stricter than the lower, as we have neglected mixing to more excited states. 

\subsection{Possible caveats of the model \label{C4:caveats}}
We have so far ignored the possibility of mixing to states made up of more than two clusters, e.g. one bound state and two free particles.  For the case of $N=4$, the possible energies of such states are much larger than the 2,2 or 1,3 geometries in the limit of strong interactions
\begin{gather}
E_{2,1,1}(X_{1,1},X_{1,2},X_{2})  =  \left(-\frac{g^2}{4}  + \frac{3}{2} +  X_{2}^2 + \frac{X_{1,1}^2+X_{1,2}^2}{2} \right)\;, \nonumber \\
E_{1,1,1,1}(X_{1,1},X_{1,2},X_{1,3},X_{1,4})  =  \left(2 + \sum_{k=1}^4 \frac{X_{1,k}^2}{2} \right) \; .
\end{gather}
These states are energetically accessible if the initial kinetic energy is larger than the interaction energy, hence initial conditions satisfying $g^2+1/2 > 4X_2^2$ should immediately see a suppressed mixing into these states.  This is however not the case if $N>4$, even with $N=6$ the $\{4,1,1\}$ state still has lower internal energy than the $\{3,3\}$ state.  

However, if the internal length scales within the bound states are much smaller than a harmonic oscillator length (hence the collision is similar to one in free space), we still expect approximate conservation of the variance in the individual momenta during a collision.  This would suppress mixing to states with, for example, all the particles sitting at the trap center. If we attempted to extend our model of oscillating clusters with Gaussian center-of-mass profiles to include more than two cluster states, the values of the outer positions $X_{n,k}$ would be much less constrained. The center-of-mass condition implies only that the sum of all the maximum positions vanishes, i.e.~$\sum_{k,n} n X_{n,k}=0$, which would have a range of solutions, some of which would be strongly mixed into during collisions and others weakly. Due to the reduced constraints, these states are very difficult to include in the model. We predict the effect on the system will be similar to that of increased relative position uncertainty between the clusters.  

In addition to this, so far we have assumed the shape of the center-of-mass wave function of each particle to be the ground state of the corresponding center-of-mass Hamiltonian. If this was allowed to vary there would be freedom to transfer kinetic energy goes into exciting this mode.  This increases the separation uncertainty between the states until the process of a collision is happening to some extent more or less continuously.  In addition to this, it is possible the clusters will go out of sync if the pseudo periodicity effect is different for each $n$, leading to further uncertainty in the relative separation.

\section{Mixing between different number configurations via time-dependent perturbation theory \label{C4:singlet_trimer_mix}}
\subsection{General setup for $N=4$}
We construct a time-dependent perturbation theory to model the transfer of population between states within our coherent state model.  Within the assumptions made we have only $N/2$ possible (time dependent) states to include.  Focusing on the case of $N=4$, we have a wave function at time $t$ of
\begin{align} 
 \psi(t) \simeq c_{2,2}(t)\varphi_{2,2}(t) + c_{1,3}(t)\varphi_{1,3}(t) \; ,
\end{align} 
which must solve the Schr\"{o}dinger equation
\begin{align}
i \frac{d}{dt} \psi(t) = [(H_{L} + H_{R}) + H_{I}] \psi(t)  \;.
\end{align}
Taking the Hamiltonian on the fundamental region $x_1 \le x_2 \le x_3 \le x_4$, and noting the time dependence of Eq.~\eqref{eq:time_evolution}, this implies
\begin{multline}
  i \left[ \dot{c}_{2,2}(t) \varphi_{2,2}(t) + \dot{c}_{3,1}(t) \varphi_{3,1}(t)  
   c_{2,2}(t) \dot{\varphi}_{2,2}(t) + c_{3,1}(t) \dot{\varphi}_{3,1}(t) \right] \\
   = 
  c_{2,2}(t) \left\{ i\dot{\varphi}_{2,2}(t) + H_I(2) \varphi_{2,2}(t)\right\} 
  \\+ c_{3,1}(t)\left\{  i\dot{\varphi}_{3,1}(t) 
    \mathcal{N}_{1,3}\left[ H_I(1) \psi_{1,3}(t)  + H_I(3)  \psi_{3,1}(t) \right]\right\}  \;,
  \label{eq:shro_pert_thr}
\end{multline}
with $H_I(n)$ given in Eq.~\eqref{eq:split_ham_2} and the dots denoting time derivatives.  Canceling terms and using the orthonormality of $\varphi_{N,n-n}(t)$, we can simplify Eq.~\eqref{eq:shro_pert_thr} to the two coupled differential equations
\begin{subequations} \label{eq:coupled_coefficients}
\begin{align}
i\dot{c}_{2,2}(t) = & \mathcal{N}_{1,3}(t)  c_{3,1}(t) \bra{\varphi_{2,2}(t)} \left[  H_I(1)  \ket{\psi_{1,3}(t)} +  H_I(3) \ket{\psi_{3,1}(t)} \right] \nonumber \\
&+ c_{2,2}(t) \bra{\varphi_{2,2}(t)}  H_I(2) \ket{\varphi_{2,2}(t)}  \\
i\dot{c}_{3,1}(t) = &\mathcal{N}_{1,3}(t)  c_{3,1}(t) \bra{\varphi_{3,1}(t)} \left[   H_I(1)  \ket{\psi_{1,3}(t)} +  H_I(3) \ket{\psi_{3,1}(t)} \right]  \nonumber \\
&+  c_{2,2}(t) \bra{\varphi_{3,1}(t)} H_I(2) \ket{\varphi_{2,2}(t)}  \;,
\end{align}
\end{subequations}
where we have introduced the compact notation
\begin{align}
 &\langle f(x_1,\ldots,x_N) |O(x_1,\ldots,x_N)| f'(x_1,\ldots,x_N)\rangle = N! \int_{-\infty}^{x_2} dx_1  \nonumber\\
 &\ldots\int_{-\infty}^{\infty} dx_N f^*(x_1,\ldots,x_N) O(x_1,\ldots,x_N) f'(x_1,\ldots,x_N) \;.
\end{align}
So far this is essentially an exact description of the two state system.  As we initially have $c_{2,2}= 1$, we can consider a perturbative solution with $|c_{3,1}(t)| \ll |c_{2,2}(t)|$ as a regime of validity. 

Formally, we perturb $[H_L(2) + H_R(2)]$ by $H_{I}(2)$. Within first order perturbation theory, this is equivalent to solving Eq.~\eqref{eq:coupled_coefficients} after dropping all terms with a prefactor of $c_{3,1}(t)$.  Hence we have
\begin{subequations}
\begin{align}
i \dot{c}_{2,2}(t) &\simeq c_{2,2}(t) \langle\varphi_{2,2}(t)|H_{I}|\varphi_{2,2}(t)\rangle, \\
i \dot{c}_{3,1}(t) &\simeq c_{2,2}(t)\langle\varphi_{3,1}(t)|H_{I}|\varphi_{2,2}(t)\rangle,
\end{align}
\end{subequations}
where we have dropped the argument of $H_{I}(2)$ for notational simplicity. We note that $\langle\varphi_{2,2}(t)|H_{I}|\varphi_{2,2}(t)\rangle$ is periodic with a periodicity $(T=\pi)$, half that of the oscillator period. The matrix element $\langle\varphi_{3,1}(t)|H_{I}|\varphi_{2,2}(t)\rangle$ is a product of a function with period $T=\pi$, and the complex exponential $\exp(-i\Delta E_{\rm int} t)$ of the energy difference between the intra-cluster degrees of freedom in both configurations $\Delta E_{\rm int} = 2E^{(2)}-E^{(3)}$.  For the range of values we consider we can use Eq.~\eqref{eq:natomenergy} for $E^{(n)}$ and $\Delta E_{\rm int} \approx -[g^2/2 + 7/(12g^2)]$ is a good estimate.

Denoting the periodic component of the interaction terms $\langle\varphi_{n,N-n}(t)|H_{I}|\varphi_{2,2}(t)\rangle$  as $f_{n,N-n}(t)$, we must therefore solve
\begin{subequations} \label{eq:coupled_coefficients_2}
\begin{align}
i \dot{c}_{2,2}(t) &\simeq c_{2,2}(t)  f_{2,2}(t) \;, \\
i \dot{c}_{3,1}(t) &\simeq c_{2,2}(t)  f_{3,1}(t) \exp\left( -i \Delta E_{\rm int} t \right) \;,
\label{eq:time_deriv_coefficients}
\end{align}
\end{subequations}
with the boundary condition $c_{2,2}(0) =1$.   The interaction time between the two states is proportional to internal size of the states and inversely proportional to the relative velocity at collision which is proportional to $1/x_0$.

\subsection{Instantaneous interaction approximation \label{C4:delta_int_pert}}

We assume $x_0$ is not small and the coupling parameter $g$ to be large and negative, corresponding to the regime of soliton like dynamics.  In this regime, excited internal states of $\ket{\psi_{n,N-n}(t)}$ (states which are not just an $n$ and $N-n$ atom ground state undergoing simple harmonic oscillation) are energetically suppressed.  Therefore $c_{3,1}(t)$ and $c_{2,2}(t)$ really do denote occupations of the states of Eq.~\eqref{eq:possible_wf2}.  The magnitude of the interaction term, $\langle \delta(x_{n} - x_{n+1}) \rangle$, does not increase with the collision velocity, but the time for which it is significant decreases asymptotically as $1/x_0$, along with its effect on the system.  Therefore if $x_0$ is large we can treat the interaction as a delta function in time and thus approximate the periodic component as
\begin{align}
f(t) \approx \sum_{k=0}^\infty \delta(t-\pi/2 - k\pi) \;. 
\end{align}
We then assume $f_{2,2}(t) = A(g) f(t)$ and $f_{3,1}(t) = B(g) f(t)$ (which also implies $A<0$ as $g<0$).   This approximation can also be considered valid even in the limit $|g| \gg 1$ with $g<0$, provided the separation is sufficiently large.  Breakdown of the approximation is expected to occur when significant relative phase evolution between the two states occurs during collision, hence $\Delta E_{\rm int} / x_0$ is effectively our small parameter.

We can use the results for phase shifts from soliton collisions in free space~\cite{LaiHaus1989B}, to give an expression for $A(g)$ in the strongly interacting limit\footnote{In~\cite{LaiHaus1989B} this phase shift is denoted as $\theta(n_1,p_1,n_2,p_2)$.}  
\begin{equation}
 A(g) \approx -2 \sum_{j=0}^n (2 - \delta_{j,0} - \delta_{j,n}) \tan^{-1} \left(|g| \frac{N/2 - n +j}{ p_{\rm rel}}\right) . 
\end{equation}
Note that $\delta_{j,k}$ denote Kronecker delta functions and $p_{\rm rel}$ is the relative momentum (per atom) between the two solitons, which in our system is a superposition of values, and for the purposes of this perturbation theory we will simply take the average value $p_{r(2,2)} \approx 2x_0$. Our perturbation theory does not formally include the interaction between left and right sides in the trimer-singlet configuration as are assuming $c_{3,1}$ is small, hence the phase of $c_{3,1}(t)$ is set purely by the unperturbed time-evolution and $c_{2,2}(t)$.  This phase shift would only matter if we solved Eq.~\eqref{eq:coupled_coefficients} with higher order time-dependent perturbation theory. 

Using this new function to solve Eq.~\eqref{eq:time_deriv_coefficients} with $c_{22}(0) =1$ gives:\footnote{Note that $\int_{-\infty}^y dx \delta(x-x_0) g(x) = \Theta(x_0-y) g(x_0)$, with $\Theta(x)$ the Heaviside step function; a sum of equally spaced step functions can be expressed as a floor function.}  
\begin{align}
 c_{2,2}(t) &\simeq \exp\left[-iA(g)\int_0^t  d\widetilde{t} f(\widetilde{t})\right]
	  & =  \exp\left(-iA(g) \left\lfloor\frac{t}{\pi}+\frac{1}{2}\right\rfloor\right) \;.
\end{align}
with $\lfloor x \rfloor$ meaning round $x$ down to the next integer. We can therefore calculate the time dependence of the other component to be    
\begin{align}
 c_{3,1}(t) \simeq -i \tilde{B}(g) \sum_{k=1}^{\left\lfloor t/\pi - 1/2 \right\rfloor} \exp\left[-ik \left(\Delta E_{\rm int} \pi + Ag\right) \right]   \;,
 \label{eq:c31}
\end{align}
with $\tilde{B}(g) = B {\rm e}^{-i\Delta E_{\rm int}\pi/2} (1 + {\rm e}^{-i Ag})/2$ a rescaled constant (an empty sum is assumed to be zero).  If all the terms in the sum of Eq.~\eqref{eq:c31} are of the same phase, within first order perturbation theory the population of the $c_{3,1}$ state will increase linearly, in what we can conjecture to be a resonant transfer.  Due to the assumption that $\vert c_{3,1} \vert^2 \ll \vert c_{2,2} \vert^2$, we cannot predict longer term trends, such as whether the population of the $c_{3,1}$ state will rise to a steady value where it will stay or if a long timescale oscillatory behavior will occur between $c_{2,2}$ and $c_{3,1}$.  We will see later in the numerics of Sec.~\ref{sec:results} that the latter effect of long timescale cyclic transfer occurs.

In order to derive an accurate expression we have found it necessary to also include a correction to the pseudo period $\delta= T_{\rm pseudo}-\pi \approx \alpha g$ (which is observed in our results, c.f.~Sec.~\ref{sec:results}) and predict the $k$th resonance to occur when
\begin{equation}
 2 k \pi = A + \Delta E_{\rm int} (\pi+\delta) \;,
 \label{eq:res_pred_better}
\end{equation}
with $\Delta E_{\rm int} = 2E^{(2)}-[E^{(3)}+E^{(1)}]$ the energy difference between the internal degrees of freedom at $g=g_{\rm rs}$ and $k$ a positive integer.  Analytic solutions are not available to Eq.~\eqref{eq:res_pred_better}, so in general it must be solved numerically.  However, we can expand $A$ as a power series in $g/x_0$ to try and predict the first resonance, $A \sim  5g/2x_0-17g^3/96x_0^3 +{\cal O}(g/x_0)^5$. Taking $\Delta E_{\rm int}\sim -g^2/2$ and approximating $A$ to lowest order $A \sim g\tilde{A}$, we can use this expression to predict low lying resonances
\begin{align}
2 k \pi &= \tilde{A}g_{\rm rs} + \frac{g_{\rm rs}^2 \pi}{2} \nonumber \\
g_{\rm rs} &= -\frac{\tilde{A} + \sqrt{\tilde{A}^2+4 k \pi^2}}{\pi} \nonumber \\
	  &\sim -2 \sqrt{k} - \frac{\tilde{A}}{\pi}\left(1 + \frac{\tilde{A}}{4 \sqrt{k} \pi} \right) - {\cal O}\left( \frac{A^4}{k} \right) \;.
	  \label{eq:res_prediction_first}
\end{align} 
Additionally we can see that in the limit $g/x_0 \to -\infty$, the phase shift $A(g) \to -3\pi$. 
\begin{align}
g_{\rm rs} \approx \frac{2\sqrt{2k+1}}{\sqrt{2}-\alpha\sqrt{2k+1}} \sim 2\sqrt{k+1/2} + \alpha(k+1/2)\;,
	  \label{eq:res_prediction_higher}
\end{align} 
In either case the proportionality to $2\sqrt{k}$ seems to be a common feature, with only small corrections from the interaction phase shift.  
We expect the predictions of Eq.~\eqref{eq:res_pred_better} to be only approximately correct and only valid in the regime when $c_{13}(t)$ is small and when the relative phase between the two states alters by an amount much less than $2 \pi$ during the collisions.  In App.~\ref{App:N>4} we extend these predictions to states with more than four atoms, keeping within the two state model.

\section{Numerical methods \label{sec:numerical}}
\subsection{Determination of the wave function and convergence}
We use an exact diagonalization method using a Fock space of Hermite functions which are eigenfunctions of the non-interacting Hamiltonian, i.e.~states of the form $\ket{n_0,\ldots,n_{\rm max}}$ with $\sum_k n_k = 4$, truncated via a bare energy cut off condition $\sum_k k n_k \le \eta$.  In order to reduce the required basis size we make use of the separability of the center-of-mass Hamiltonian as outlined in our previous work~\cite{HoldawayWeiss2012}; this projects the matrix elements of operator expectation values into a basis in which the center-of-mass is in the ground state, considerably reducing the basis size. 

\begin{figure}[ht!]
\begin{center}
\includegraphics[width=\linewidth]{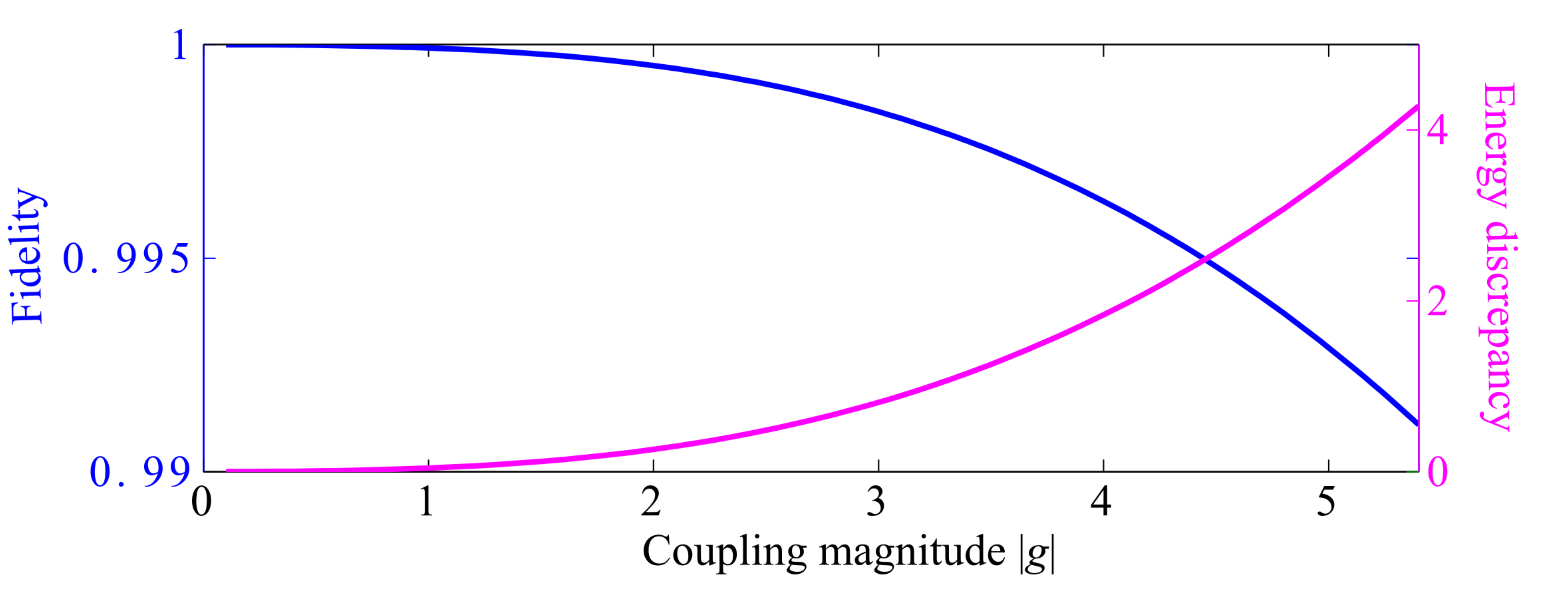}
\caption{Color online: Fidelity [left axis, dark line, defined via Eq.~\eqref{eq:fidelity}] and energy difference [right axis, light line, defined via Eq.~\eqref{eq:Ediscrep}] between the numerical initial condition and analytic initial condition for a range of $|g|$ (with $g<0$).  Initial energy is always overestimated in the numerics as the basis set cannot accurately resolve states on the length scales required.  When considering the relative size of the energy discrepancy, we note that the interaction energy for our initial condition scales as $-g^{2}$ in the harmonic units used.}
\label{fig:numerical_issues}
\end{center}
\end{figure}

Despite this reduction in basis size, the need to resolve two length scales (the trap of order unity and internal structure of the solitons scaling as $1/|g|$) means that for a finite basis size this method will always fail for sufficiently large and negative $|g|$ values.  We quantify this discrepancy via two quantities, the fidelity with our numerical initial state $\psi_{\rm num}(x_{1},\ldots,x_{N},0)$ and the actual initial state $\psi(x_{1},\ldots,x_{N},0)$:
\begin{equation}
 {\rm Fidelity} = \vert\bket{\psi_{\rm num}(x_{1},\ldots,x_{N},0)}{\psi(x_{1},\ldots,x_{N},0)} \vert^2 \;,
 \label{eq:fidelity}
\end{equation}
and the energy difference between the two (constant for all time):
\begin{align}
 \mbox{Energy discrepancy} = &\bra{\psi_{\rm num}(x_{1},\ldots,t)} H \ket{\psi_{\rm num}(x_{1},\ldots,t)} \nonumber \\
 &- \bra{\psi(x_{1},\ldots,t)} H \ket{\psi(x_{1},\ldots,t)}\;,
 \label{eq:Ediscrep}
\end{align}
with $H$ the full Hamiltonian.  This is plotted on Fig.~\ref{fig:numerical_issues}, with $x_0=3$ and a cut off $\eta =113$.  While the fidelity remains good $(>0.99)$ over the range we considered, the energy difference grows to several harmonic oscillator units.  The effect of this energy difference is discussed later in the results section.  Raising $\eta$ improves convergence, however there are technical difficulties limiting our ability to do this.  Although the final number of basis states (in the reduced basis) remains manageable, the projection of the Hamiltonian and matrix representations of operators into this reduced basis involves calculating all the matrix elements in the full basis, before projecting into the zero center-of-mass excitation basis.  This full basis scales much more rapidly as $\eta$ is increased, and so initially computing all the operators becomes increasingly demanding numerically.

\subsection{Operator of interest}
Time evolution of the wave function is easily obtained once the Hamiltonian is diagonalized, by projecting into occupations of the numerical eigenstates.  Analytically we are primarily interested in the populations of the $\varphi_{3,1}(t)$ and $\varphi_{2,2}(t)$ states, however due to many neglected effects it is not practical to look for these directly.  We instead investigate properties related to the number-to-the-right operator
\begin{equation}
 \hat{N}_{R} = \int_0^{\infty} dx \hat{\Psi}^{\dagger}(x) \hat{\Psi}(x) \; ;
\end{equation}
where $\hat{\Psi}(x)$ is the Bosonic field operator. This operator is particularly interesting as it is possible to measure this quantity experimentally in the scheme described in Sec.~\ref{sec:init_cond} and~\cite{HoldawayWeiss2013}, one could switch on the second laser to return to a deep, double frequency super-lattice and directly image each lattice site. Firstly we consider the standard deviation of its expectation value 
\begin{equation}
\Delta N_R = \sqrt{\ev{\hat{N}_{R}^2} - \ev{\hat{N}_{R}}^2} \; ,
\label{eq:DNR}
\end{equation}
noting that by symmetry $\ev{\hat{N}_{R}}=N/2=2$.  Additionally we are interesting in looking at the positions of the oscillating dimer, singlet and trimer states.  In order to achieve this we consider splitting our wave function into eigenstates of $\hat{N}_{R}$, denoted $\ket{\psi_n}$, which satisfy $\hat{N}_{R} \ket{\psi_n} = n \ket{\psi_n}$, hence we have 
\begin{equation}
 \ket{\psi} = a_0\ket{\psi_0} +  a_1\ket{\psi_1} + a_2 \ket{\psi_2} +  a_3\ket{\psi_3}  + a_4 \ket{\psi_4} \; .
 \label{eq:numberdecomp}
\end{equation}
By symmetry we must also have $a_0=a_4$ and $a_1=a_3$, two are redundant and define
\begin{subequations}
 \begin{align}
 P_{n,N-n}(t) &= |a_n|^2 + (1-\delta_{n,N/2}) |a_{N-n}|^2  \\
 R_{n,N-n}(t) &= \bra{\psi_n} x \Theta(x) \ket{\psi_n} \\
  \sigma_{n,N-n}(t) &= \sqrt{\bra{\psi_n} x^2 \Theta(x) \ket{\psi_n} - R_{n,N-n}(t)^2} \;,
\end{align}
    \label{eq:num_resolved}
  \end{subequations}
with $\Theta(x)$ the Heaviside step function. These quantities are the probability to find $n$ or $n-N/2$ atoms to the right (and $n-N/2$ or $n$ to the left) of the trap, the right side expectation value for a given number of atoms and the standard deviation of this quantity respectively.  

\subsection{Relationship with the two state model}

In order to relate these values to our two state model described in Sec.~\ref{sec:cohstatemod}, we note that when there is little overlap between the left and right hand sides $P_{4,0}(t)$ should be close to zero and if we additionally assume there is not a significant population of states with three or more bound states we can say $P_{n,N-n}(t) \approx |c_{n,N-n}|^2$. When the left and right states overlap during collisions, it is no longer possible to distinguish the left and right sides and there is a higher probability to find all four atoms to one side of the trap regardless of the relative populations $c_{1,3}$ and $c_{2,2}$.  $P_{4,0}(t)$ can reach a maximum of $1/2^4$, the value it would take for a product state located at the trap centre. Therefore $P_{4,0}(t)$ can be used to quantify the amount of collision occurring, an increase in position uncertainty between left and right states will cause this value to be less peaked as collision time becomes less well defined as was the case in~\cite{HoldawayWeiss2013}.  

The quantity $\Delta N_R$ is related in much the same way to the coefficients $c_{1,3}$ and $c_{2,2}$, again only in the absence of any left right overlap.  In terms of the number state decompositions of Eq.~\eqref{eq:numberdecomp} we have $\Delta N_R = \sqrt{ \vert a_1\vert^2+2^2 \vert a_2\vert^2 + N^2\ldots +\vert a_N\vert^2 - (N/2)^2}$.  If we consider again that $P_{4,0}(t) = 2|a_4|^2 \approx 0$, and that $|a_2|^2 + 2|a_3|^2 + 2|a_4|^2= 1$, we can simplify to obtain $\Delta N_R^2 \approx 10 |a_3|^2 + 4 |a_2|^2 - 4 = 2|a_3|^2$ which is in turn equal to $|c_{1,3}|^2$ within the time dependent perturbation theory. Therefore it would seem a value of $\Delta N_R = 1$ while $P_{4,0}(t)$ remains negligible would imply complete population transfer to the triplet singlet state has occurred. However this is complicated by the fact that the configurations with more than two bound states, discussed in Sec.~\ref{C4:caveats}, may also be populated.  We note that we have no direct way to examine this, which will remain an issue in interpreting the results and will be an area for future research.

\section{Results \label{sec:results}}
\subsection{Effect of initial position $x_0$}
\begin{figure}[ht!]
\begin{center}
\includegraphics[width=\linewidth]{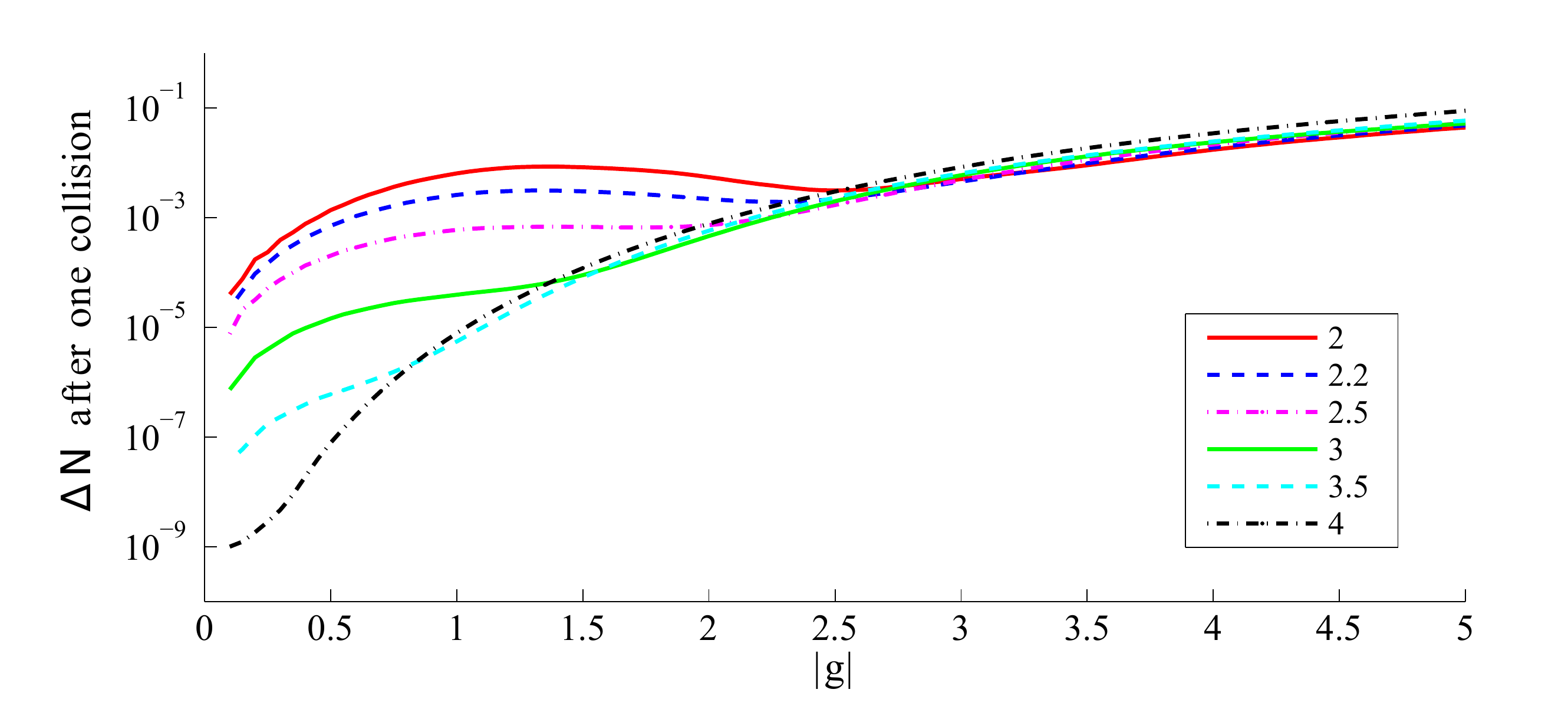}
\caption{Color online: Minimum left-right number uncertainty $\Delta N_R$ after a single collisions for different values of $x_0$ (given in the legend). Note that the values at $g=0$ should be of the order $\exp(-x_0^2)$.  For large $|g|$ the relationship is quadratic.}
\label{fig:x0_var_transfer}
\end{center}
\end{figure}  

Figure~\ref{fig:x0_var_transfer} shows the (minimum) value of $\Delta N_R$, given by Eq.~\eqref{eq:DNR}, obtained after the first collision.  Note that even as $g \to 0$ this quantity is finite and of the order $\exp(-x_0^2)$ from the finite overlap at the trap center in the initial condition, and is therefore not directly related to the trimer-singlet state population, but should give a good estimate.  Two clear regimes emerge, the low $|g|$ regime in which lower $x_0$ leads to more transfer due to the additional interaction time, and a high $|g|$ regime in which $\Delta N_R$ increases with high $x_0$.  The latter may be explained by the additional energy allowing for transfer to three or four cluster states, and/or that the reduced interaction time means that the dimer-dimer state evolves less in phase during the collision, leading to less destructive mixing into the trimer-singlet state.

\subsection{Identification of an effective pseudo period}
\begin{figure}[ht!]
\begin{center}
\includegraphics[width=\linewidth]{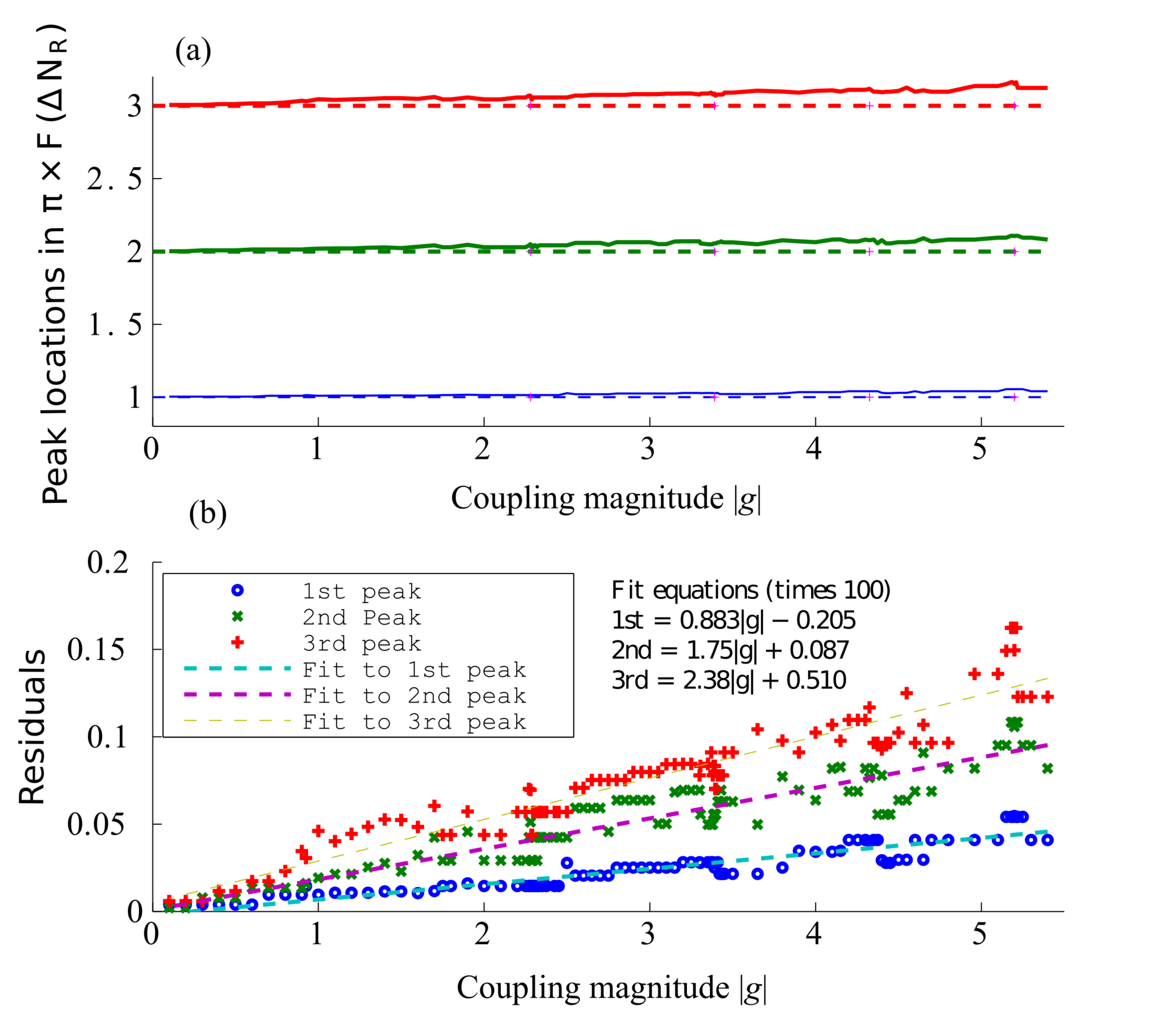}
\caption{Color online: With $x_0=3$: (a) Frequency of the first three peaks in the Fourier transform of $\Delta N_R(t)$ times $\pi$ compared to the non interacting values; pluses in the dotted line denote approximate locations of the transfer resonances. (b) Difference between these peak locations and the non-interacting values along with linear fits, equations for fits are shown in the figure inset.}
\label{fig:pperiod}
\end{center}
\end{figure}  

Attractive interactions between the left and right bound state clusters are expected to shorten the effective period of oscillation.   
Figure \ref{fig:pperiod} (a) shows the peaks in the Fourier transform of $\Delta N_R(t)$ at different values of $|g|$ (with $g<0$) along with the values at $g=0$, (b) shows the residuals between these lines, along with linear fits.  The constant offsets should be zero, however the data is not perfectly linear and is not expected to be. These fits can be used to identify an effective pseudo-period, we are interested in the first line (the other two are approximately integer multiples of the gradient) which can be used to improve our predictions of the resonances. Denoting this first fit as $\tilde{\delta} \approx -0.009g$, this term is related to \protect{$\delta = T_{\rm pseudo}-\pi$} in Sec.~\ref{C4:delta_int_pert} via $\delta = -\tilde{\delta} \pi/(1+\tilde{\delta})$.  

\subsection{Position of the transfer resonances}

\begin{figure}[ht!]
\begin{center}
\includegraphics[width=\linewidth]{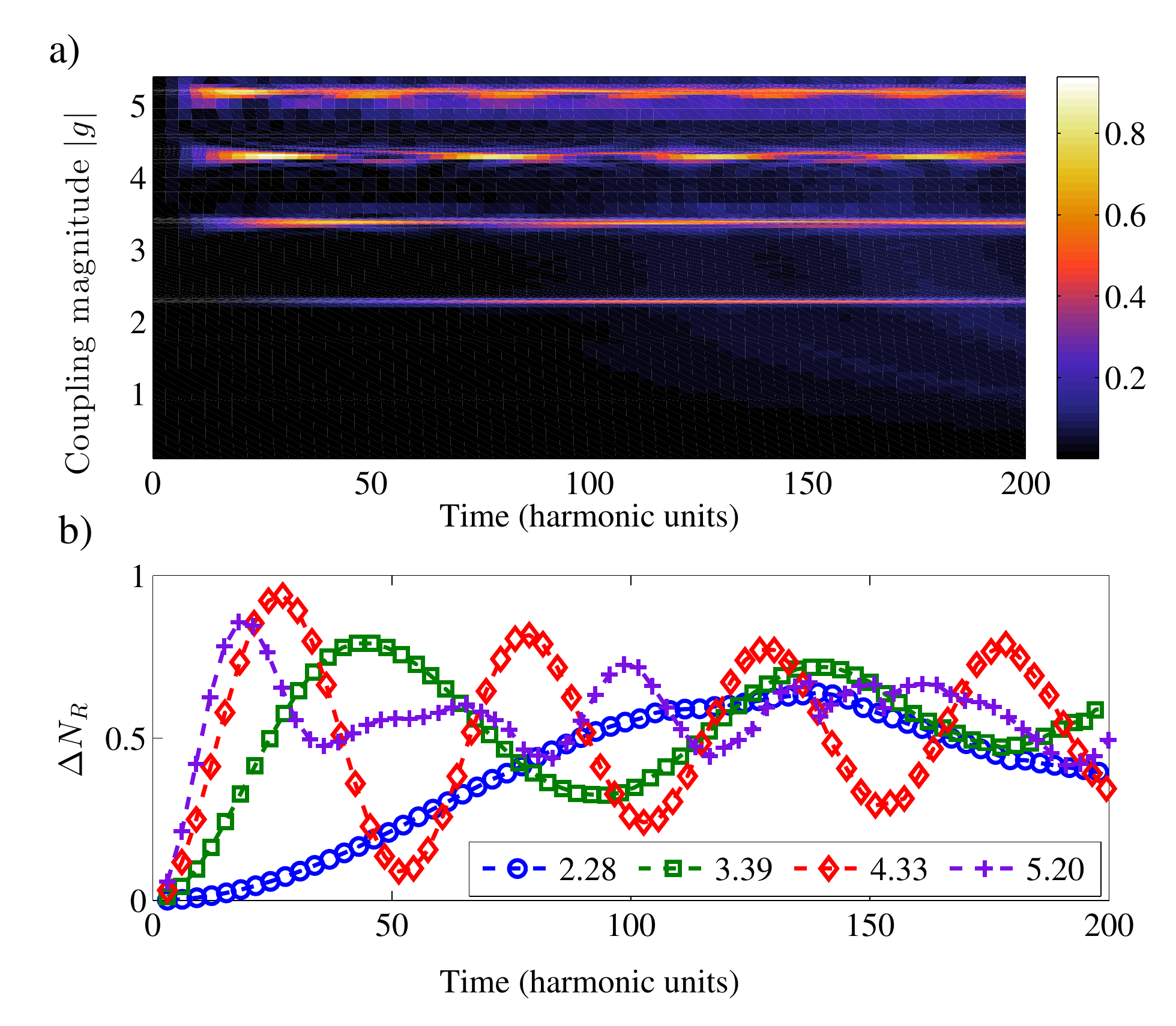}
\caption{Color online: (a) Minimum value taken by $\Delta N_R$ after a given collision for a range of $|g|$ (with $g<0$). (b) Lines corresponding to near resonant values of $|g|$. The oscillatory behavior appears to occur with shorter periods at higher $|g|$, the largest value however does not follow this pattern and is likely an artifact from the numerical breakdown.}
\label{fig:res_comp}
\end{center}
\end{figure}  

In Sec.~\ref{C4:delta_int_pert} we derived analytic estimates for the values of $g$ at which the transfer resonances were expected to occur.  By numerically solving Eq.~\eqref{eq:res_pred_better} (with $\delta$ obtained from Fig.~\ref{fig:pperiod} as discussed in the period section) for $x_0=3$ we estimate the first four resonances to be at 
\begin{equation}
 g_{\rm rs} \sim -2.4,\; -3.2,\; -3.9,\; -4.4 \;.
\end{equation}
Our numerical calculations, plotted in Fig.~\ref{fig:res_comp} (a) show the minimum value obtained by $\Delta N_R$ after a particular collision (data is in discrete blocks around $\pi$ in time).  For particular values of $g$ this reaches large values to a phase matching condition leading to constructive mixing between the two states.  
At first the numerics agree with our prediction, with the first resonance observed at $g=-2.28$, however the higher predicted values agree progressively less and less well with the observed positions, being found at $g \sim -3.4,-4.3,-5.2$.  The reason for this is almost certainly that the finite basis set used means that $(E_{2,2}-E_{3,1})$ is inaccurate.  We expect the numerically calculated energy difference to be lower,\footnote{While both $E_{2,2}$ and $E_{3,1}$ will be overestimated in the numerics, it is possible the difference could be larger if $E_{2,2}$ is overestimated much more than $E_{3,1}$.} than the analytic estimates.  This has the effect of shifting the resonances to higher values and our numerical results are qualitative rather than quantitative.  We note that neither the analytic nor numerical resonance position values shift significantly as $x_0$ is varied.  

There are also the issues of the impact of confinement, which should make the solitons narrower for a given $g$,  and therefore modify phase shift from collisions.  Additionally there is a distribution of momenta which has not been accounted for, we have simply used the mean value.  Finally the back-reaction of the terms proportional to $c_{3,1}$ in Eq.~\eqref{eq:coupled_coefficients} could shift the resonances by an amount of the order of the width of the resonance, and the finite interaction time becomes more important for increasing $|g|$.  However, none of these effects are expected to be significant enough to account for the discrepancy at large $g$, which is currently a numerical issue. 

Figure \ref{fig:res_comp}(b) takes values close to these resonances, the fourth resonances (`+' markers) does not follow the same long time oscillatory pattern as the others, which we assume to be due to numerical breakdown.  In principle the values of $\Delta N_R$ when the states are well separated should be equal to $c_{3,1}$ in our two state model.  However, an increase in relative position uncertainty between the left and right states means that some part of the wave function is always in collision and hence the oscillations appear to become less extreme at later times. This effect is explained in more detail in the next section. 

An important experimental consideration is the effect of decoherence.  Interactions with the wider environment could act in similar way to a number measurement on one side of the system, transforming the state from a superposition of three left and one right and vice versa to a statistical mixture of both possibilities.  The state of three atoms on one side and one on the other is however not a fully symmetric state about the trap center. The odd components of the wave function cannot mix back to the original state, which will further reduce the partial revivals in Fig. \ref{fig:res_comp}(b).  These revivals could be experimentally measured by repeated runs of an experiment, and performing number measurements at different times.  This would provide strong evidence for the quantum mechanical nature of the superposition (or lack thereof).

\subsection{Number resolved dynamics}

\begin{figure}[ht!]
\begin{center}
\includegraphics[width=\linewidth]{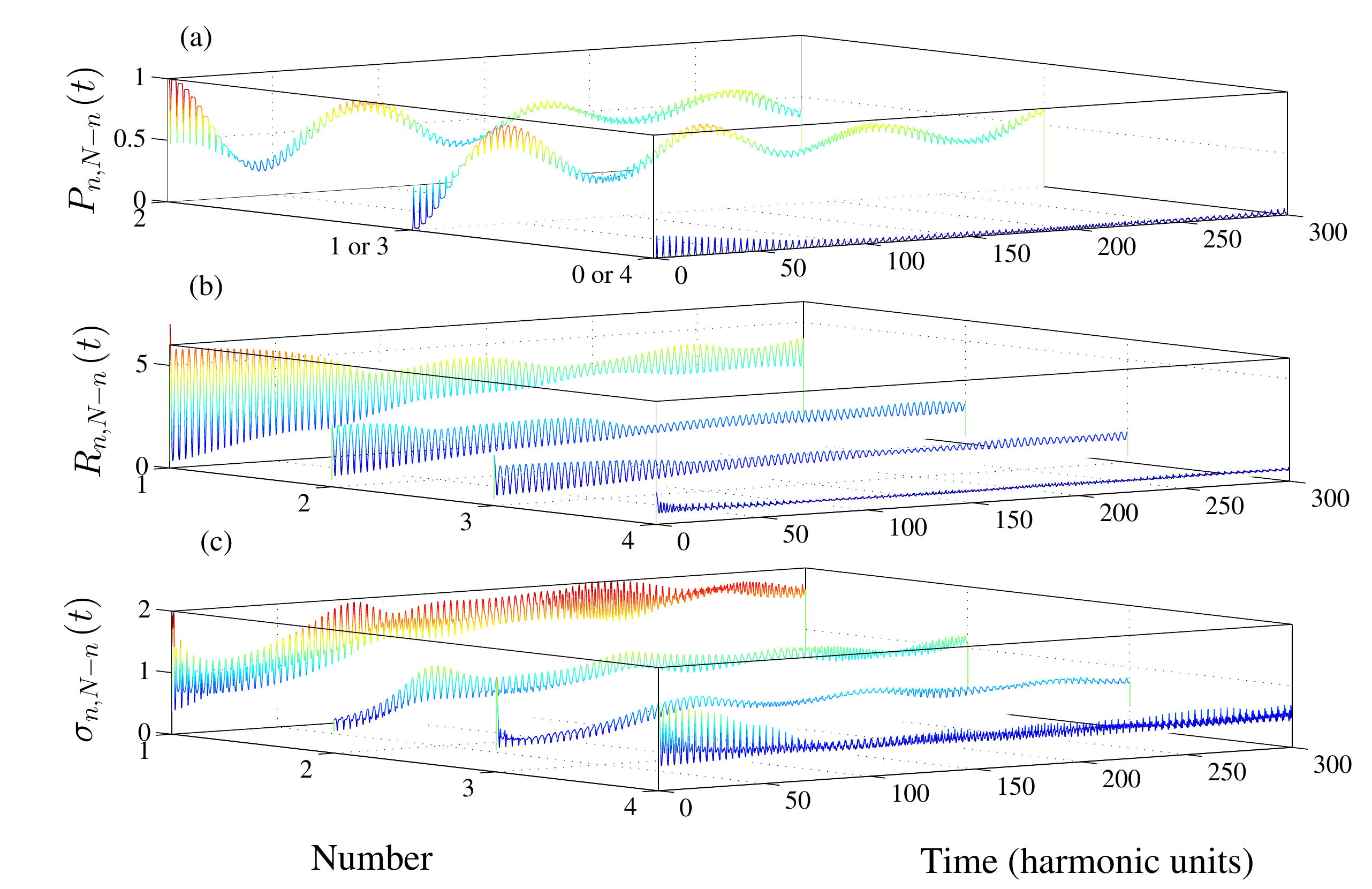}
\caption{Color online: (a) Probability of finding $n$ atoms to the right of the trap, (b) shows right side position expectation value on components of the wave function with exactly $n$ atoms to the right, and (c) shows the variance in this value.}
\label{fig:waterfall}
\end{center}
\end{figure}  

We now focus on the quantities defined in Eq.~\eqref{eq:num_resolved}, these focus on amplitudes of different number states and expectation values of right-side operators.  For again $x_0 = 3$ and $g=-3.39$ (second resonant value), Fig.~\ref{fig:waterfall}(a) shows how these amplitudes, relating to the probability of a measurement of $\hat{N}_R$ gives $0$ or $4$, $3$ or $1$ and $2$, vary in time.  $P_{4,0}$ should be negligible except during collisions, this is initially the case but gradually appears to ``smear out'', with less well defined peaks.  This effect is quite significant, even by $t=50$ when maximum transfer is has occurred to the trimer-singlet state at $g=-3.39$ and so we cannot determine exactly how much population has been transferred.  

Figure~\ref{fig:waterfall}(b) shows the $R_{n,N-n}$ values, which follow particle like tracks in a harmonic oscillator potential, except where they near zero and this quantity is less meaningful. The maximum amplitude of the single particle oscillation is $5.9$ whereas Eq.~\eqref{eq:E1=E2} predicts an amplitude of $5.6$, in reasonable agreement given the assumptions made in the model and limited numerical convergence of the state energies.\footnote{The total energy for these parameters is around 1 harmonic energy unit great than would analytically be predicted.} The fact that the early time expectation values for $n=3$ and $n=1$ roughly follow these harmonic oscillator tracks, as predicted within the two state model, combined with the increasing probability $P_{3,1}$, is the strongest evidence for the importance of the trimer and single atom superposition state.  If the collisions were mixing significant population into states with more than two clusters, then $R_{1,3}$ could not obtain such a high maximum value because of energy conservation.     The position uncertainty increase means that at late times these expectation values no longer have the same clearly defined particle-like tracks, tending more towards the time averaged displacement in the particle-like trajectories.  Finally~\ref{fig:waterfall}(c) shows the right-side-position uncertainty of each of the number-resolved values, which increases to a maximum as the relative position uncertainty between the left-and-right states becomes larger for all possible number configurations.  This is further evidence for an increase in the uncertainty in the separation between the left and right sides.

\section{Conclusions and outlook \label{sec:conc}}
We have derived an analytical model that predicts collisions between quantum solitons (initially of definite number) in the presence of harmonic confinement will introduce mixing into states with different numbers of atoms in each soliton, thus creating a relative number uncertainty between the soliton.  This model also allows us to predict particle like trajectories for each different number state oscillating in the trap, although it has not been possible to analytically include the increased relative position uncertainty from collisions in our current work.

Our simple two cluster model predicts that for specific values of the interaction strength, $g$, a phase matching condition is achieved between initial state of two $N/2$ clusters and a state which is a superposition of $N - 1$ atoms to the left (right), and one atom to the right (left).  This means repeated collisions transfer populate to this state (and possibly others) constructively, at least while the populations of this state is much lower than the population of the state with two $N/2$ clusters, leading to what we refer to as a transfer resonance.  

We observe these transfer resonances in the numerics for $N=4$, cycling population between the dimer-dimer and trimer-singlet states on long time-scales.  Stronger interactions make this transfer faster, allowing for a state of very high relative number uncertainty to be created before the increase in relative position uncertainty means it is no longer possible to determine at what times the states are well separated.  However our numerical method, based on exact diagonalization, suffers some convergence issues for the $|g| \gg 1$ regime, leading to discrepancies between our numerical energies and those predicted analytically.  This is thought to shift the resonances from their true locations and possibly to affect the rate of transfer, however it confirms that such resonances should exist, at least for the $N=4$ case. Additionally the numerics indicate that the collisions transfer populations between the states cyclically on long timescales, beyond what is possible to predict from first order perturbation theory.  Despite the convergence issues, we still expect the numerics for the lowest resonance to be accurate as the energy and fidelity discrepancies between our analytical initial condition and our numerical initial condition (energy is conserved throughout the simulation) are small, and the second and third resonances should be qualitatively correct. 

It would be of great interest in future research to test the general predictions made via our perturbation theory for larger numbers of atoms, shown in appendix~\ref{App:N>4} .  It may be possible to use our numerical method with larger numbers of atoms, however other more adaptive methods such as TEBD~\cite{Vidal2004,Daley2013} and MCTDHMB~\cite{AlonCederbaum2008} are also available to study many body dynamics if this proved too difficult.  Numerics based on the Bethe ansatz states should also be a possibility in the strongly interacting regime.

\begin{acknowledgements}
We would like to thank the UK EPSRC for funding (Grants EP/G056781/1 and EP/K03250X/1).
\end{acknowledgements}

\begin{appendix}
\section{Resonant predictions for $N>4$ \label{App:N>4}}
\subsection{Predictions within the two cluster model}
We can also extend the time-dependent perturbation theory of Sec.~\ref{C4:delta_int_pert} in order to consider the situation for $N>4$, if we again assume our state space to be limited to that of two cluster states [formalized in Eq.~\eqref{eq:possible_wf2}] and proceed with the same time dependent perturbation theory, assuming all $c_{N/2-n,N/2+n}$ except $c_{N/2,N/2}$ are small.\footnote{A different model for which $n$-th order perturbation theory is produces small results for $n=1,2\ldots n'-1$, but large results for $n=n'$ can be found in~\cite{EsmannEtAl2012}.}  The relative energy difference between two clusters of $N/2$ atoms and a state with clusters of size $N/2-n$ and $N/2+n$, is given in the limit $|g| \gg 1$, $g<0$ as
\begin{equation}
 \Delta E_{\rm int}(N/2-n,N/2+n) \sim \frac{g^2 N n^2}{8} \;.
\end{equation}
We follow the same procedure as before to derive a condition for $c_{N/2-n,N/2+n}$ to increase resonantly, taking $A_{n}  \approx \theta(N/2,N/2,2x_0)$.  Because we are not going to numerically determine these resonances, we just expand $A_{n} = g\tilde{A}_{n} + {\cal O}(g^2)$ to examine the low lying resonances, which gives
\begin{align}
g_{\rm rs}(k) &= -4 \left|\frac{\tilde{A}_{n} \pm \sqrt{\tilde{A}_{n}^2+N k n^2 \pi^2}}{n^2 N\pi} \right|\nonumber \\
	  &\sim -4 \frac{\sqrt{k}}{n\sqrt{N}} - {\cal O}\left( \frac{\tilde{A}_{n}}{n^2 N} \right) \;,
	  \label{eq:res_prediction_N}
\end{align} 
for the $k$th resonance. For simplicity, we assume all $A_{n}$ to be negligible in what follows. This should always be possible in the limit of $x_0 \gg 1$, as the interaction time tends to zero.  Letting $n = N/2-1$, $k=1$, we obtain the resonant condition for mixing to the singlet plus $N-1$ atom bound-state configuration, which is achieved for the weakest interaction strength of $g \sim -4/[(N/2-1)\sqrt{N}]$.  We note that for fairly modest $N \gtrapprox 10$, this begins to violate the strong interaction condition $N|g| \gg 1$ and so would need to be modified for general $N$.  Full population transfer to this state would give $\Delta N_R^2 = N^2/4-N+1$, which has the same leading order term as NOON state, making it potentially interesting for interferometry.  Additionally, because this state is the most energetically favored, all the others are non resonant (in fact likely close to anti-resonant) and should only accumulate small population.  The reduction in the magnitude of the rescaled interaction strengths may make this more experimentally favorable.


\subsection{Effect of configurations with three of more bound states}

With $N > 4$, states with three or more clusters are not all energetically suppressed, as was the case in our discussion in Sec.~\ref{C4:caveats}; mixing to such states will likely yield new and richer physics, beyond that considered in our perturbation theory.  However, these multiple cluster states will still have an internal phase evolution which should not evolve by a multiple of $2\pi$ every half oscillator period when the $N-1,1$ state is at its first resonance, due to the fact that the internal energy of any 3 cluster state must be higher.  For instance, the state with an $N-2$ atom bound state and two free atoms, all of which are undergoing simple harmonic oscillation and considered only to interact during collisions, will have an internal energy of
\begin{multline}
 2E^{(N/2)} - E^{(N-2)} \\ \approx 
  \frac{g^2\left[(N-3)(N-2)(N-1) - (N^2/4-1)N\right]}{24} \;,
\end{multline}
in the strongly interacting limit.  As we mentioned this is not valid for $N \gtrapprox 10$ as $g = -8/[(N-2)\sqrt{N}]$ where we expect the first transfer resonance. At this point we have
\begin{align}
 \Delta E_{\rm int} = \frac{N^2-6 N+4}{4 N (N-2)} \;,
\end{align}
meaning the phase will not change by an integer fraction between collisions, and hence this state cannot be resonant.  

It therefore seems quite possible that these transfer resonances would still be present for larger $N$, at least into the $N-1$ bound state plus single atom configuration, and would be a viable method to create states with very non-classical relative number statistics about their center. However, further numerics or experiments would be needed to confirm this.  

\end{appendix}

\end{document}